\documentstyle[aps,prl,epsfig,floats,eqsecnum]{revtex}

\begin{document}
\draft
\twocolumn[
\hsize\textwidth\columnwidth\hsize\csname 
@twocolumnfalse\endcsname
\title{Stripe patterns in a granular system induced by slow deformation
of its container}
\author{So Kitsunezaki and Akemi Kurumatani}
\address{
Department of Physics, Graduate School of Human Culture, Nara Women's
University, Japan}
\date{\today}
\maketitle
\begin{abstract}
We investigate the formation of stripe patterns that appear on
 the surface of a dry granular system as the container is deformed very slowly.
In an experimental study using nearly mono-disperse glass beads, we found that many faults develop beneath the surface.
Our results show that the spacing of stripes is independent of the system
 size and does not depend significantly on the grain size. 
\end{abstract}
]

%%%%%%%%%%%%%%%%%%%%%%%%%%%%%%%%%%%%%%%%%%%%%%%%%%%%%%%%%%%%%%%%%%%%%%%%%%%
%\section{Introduction}
%%%%%%%%%%%%%%%%%%%%%%%%%%%%%%%%%%%%%%%%%%%%%%%%%%%%%%%%%%%%%%%%%%%%%%%%%%%
Granular systems under the influence of gravity exhibit a rich variety
of interesting phenomena \cite{Jaeger96,deGennes98,Makse98}.
In the solid phase, it is well known that granular systems differ
greatly from ordinary solids with regard to statistical
properties of stresses and that the stress distributions within them 
depend strongly on the history of their formation \cite{Mueth98,Vanel99,Geng00}.
Also the deformation of granular matter is usually accompanied
by sheet-like localized fluidization appearing in an otherwise solid phase.
Such a fluidized region is called a ``shear band'', and 
if it is formed in the bulk, rather than the surface, it is called a ``fault'' \cite{Lemieux00,Mueth00,Astrom00}.

We experimentally investigated the formation of stripe patterns formed
on the surface of a system of granular matter as a result of the slow
deformation of its container\cite{Kitsunezaki01a}.
We used spherical glass beads with a nearly
mono-disperse size distribution.
Our experiments show that these stripes were caused by the formation of
many faults beneath the surface, not fluidization on the surface.

Although our investigations were inspired by the accidental finding of 
one of authors(Kurumatani), we recently realized that in 1928 T. Terada
and N. Miyabe reported similar phenomena in the formation of a series of faults in dry sand \cite{Terada}.
They are pointed out similarities with certain geological phenomena.
They also found that the faults are approximately parallel and that the
angle of inclination of the faults depends on the order of packing.
The behavior of these faults reminds us of the formation and arrangement 
of shear bands in rocks, which have been investigated by several people \cite{Poliakov94}.
The mechanism responsible for the arrangement of faults in granular
systems has not yet been revealed.
Here we report experimental results elucidating the dependence on
various conditions of the spacing of stripes created by faults.  

\

%%%%%%%%%%%%%%%%%%%%%%%%%%%%%%%%%%%%%%%%%%%%%%%%%%%%%%%%%%%%%%%%%%%%%%%%%%%
%\section{Experimental Methods}
%%%%%%%%%%%%%%%%%%%%%%%%%%%%%%%%%%%%%%%%%%%%%%%%%%%%%%%%%%%%%%%%%%%%%%%%%%%
\begin{figure}[h]
\begin{center}
\epsfig{width=4.5cm,file=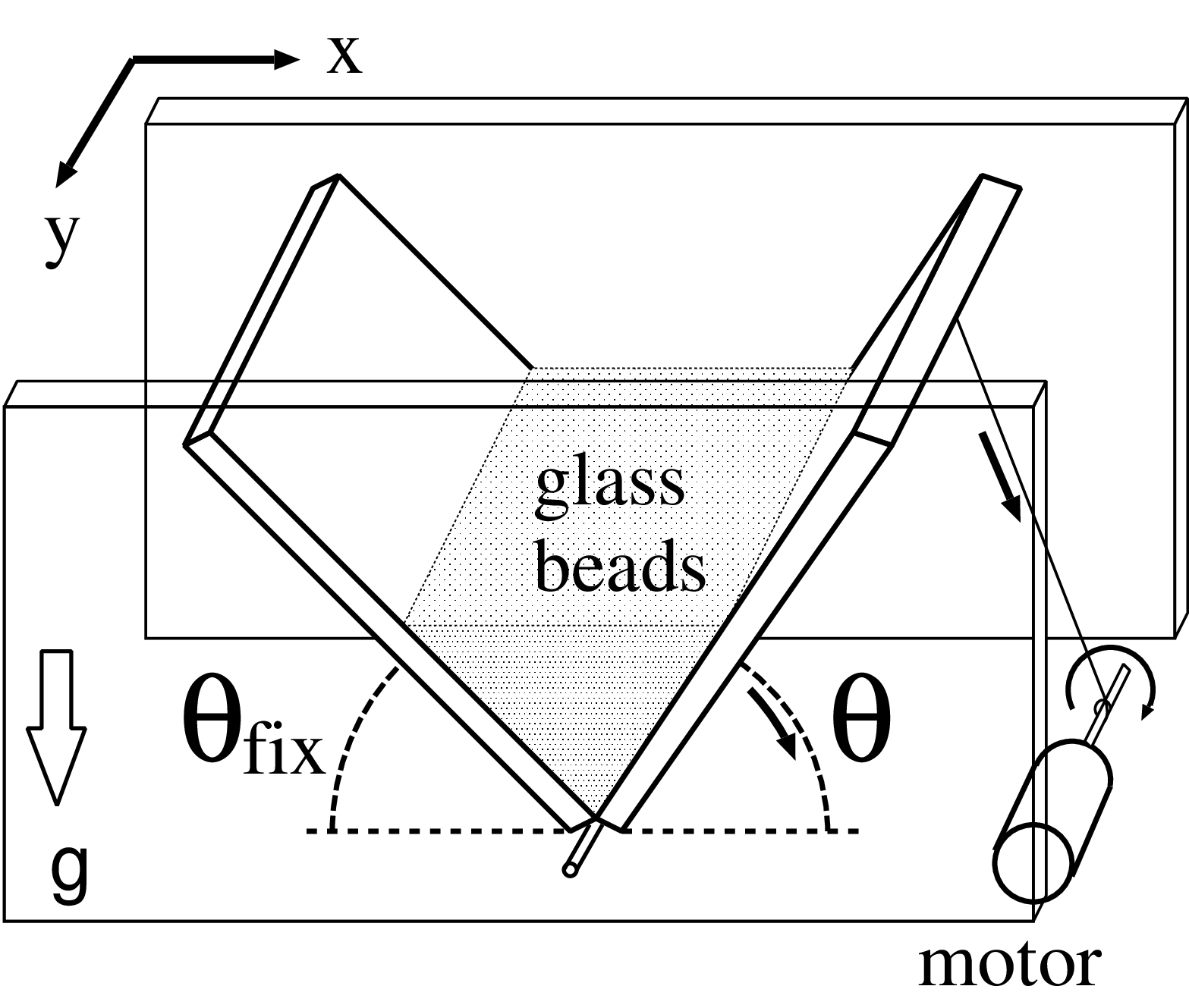}
\caption{The experimental setup.}\label{setup}
\end{center}
\end{figure}
We prepared three kinds of spherical glass beads, made of soda-lime
silicated glass of density $2.5 \pm 0.1 g/cm^3$, with average diameters 
$D=40, 100$ and $200 \mu m$.
We refer to them as GB40, GB100 and GB200.
Their size distributions were nearly mono-disperse, with $80-90\%$ of
the particles having diameters between $0.9D$ and $1.1D$ \cite{beads}.
We used a fourth kind of glass beads, with 0.1 mm diameter (GB-01), 
only for experiments in which we prepared colored layers of beads.
The size distribution of GB-01 was somewhat less mono-disperse than
those of the other beads.

A schematic depiction of the container and glass beads used in our
experiments is given in Fig.\ \ref{setup}. 
As shown there, two rectangular plates are joined to form a V-shape
structure held between parallel lateral walls.
We refer to the plates as ``V-walls''.
V-walls and the lateral walls are all made of acrylic plates and V-walls
are covered with thin rubber sheets to protect the acrylic from being scratched. 
The slope of each V-wall is variable.
In an experiment, we fixed one of the plates at an angle
$\theta_{fix}$ and decreased the angle of the other plate, $\theta$, from some initial angle $\theta_0$ at an almost constant angular velocity, $\omega$.
For this purpose, we used a motor connected to a power supply
with a constant voltage to pull this plate with a string.
We first poured a single kind of beads with total weight $M g$ into the
container and tapped it a few times to level the surface horizontally.
For each experiment, the volume fraction of glass beads in the container was $0.53-0.61$.
The volume fraction depended slightly on the type of beads used, and
it was not exactly the same for each experiment.
However, the results described below seem to be largely insensitive to
these differences. 

\begin{figure*}[t]
\begin{center}
\begin{minipage}{5cm}
\epsfig{width=5cm,file=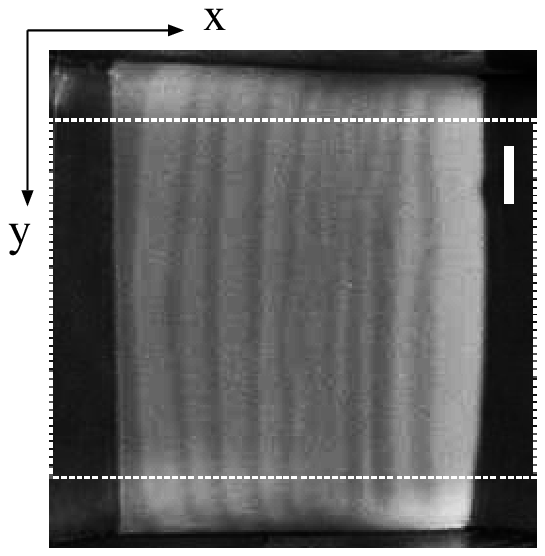}
\caption{A snapshot of a stripe pattern for $\theta=30^{\circ}$.
}\label{snapshot}
\end{minipage}\hspace{5mm}\
\begin{minipage}{5cm}
\epsfig{width=5cm,file=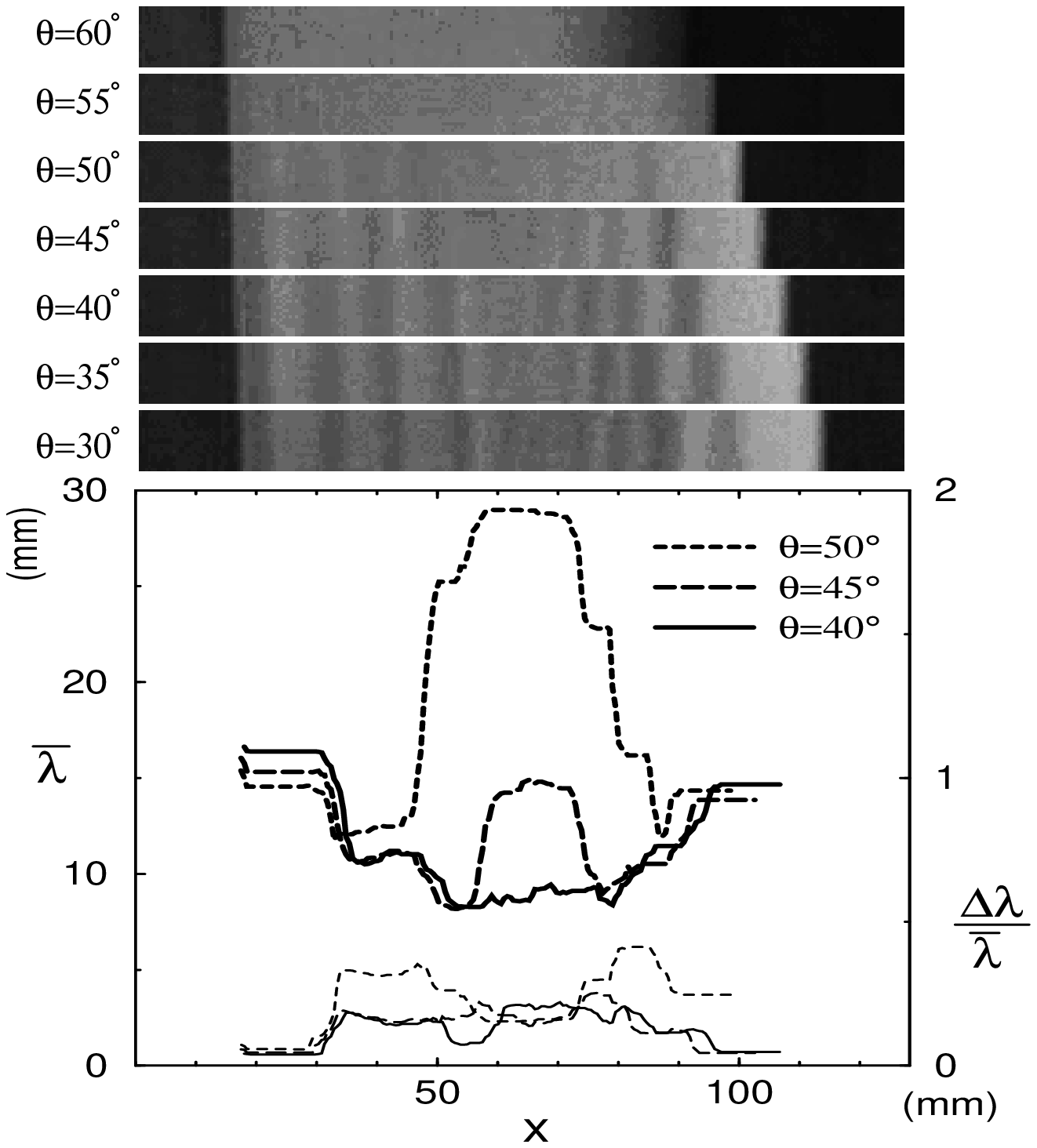}
\caption{The process of the formation of stripes.
}\label{move}
\end{minipage}\hspace{5mm}\
\begin{minipage}{5cm}
\epsfig{width=5cm,file=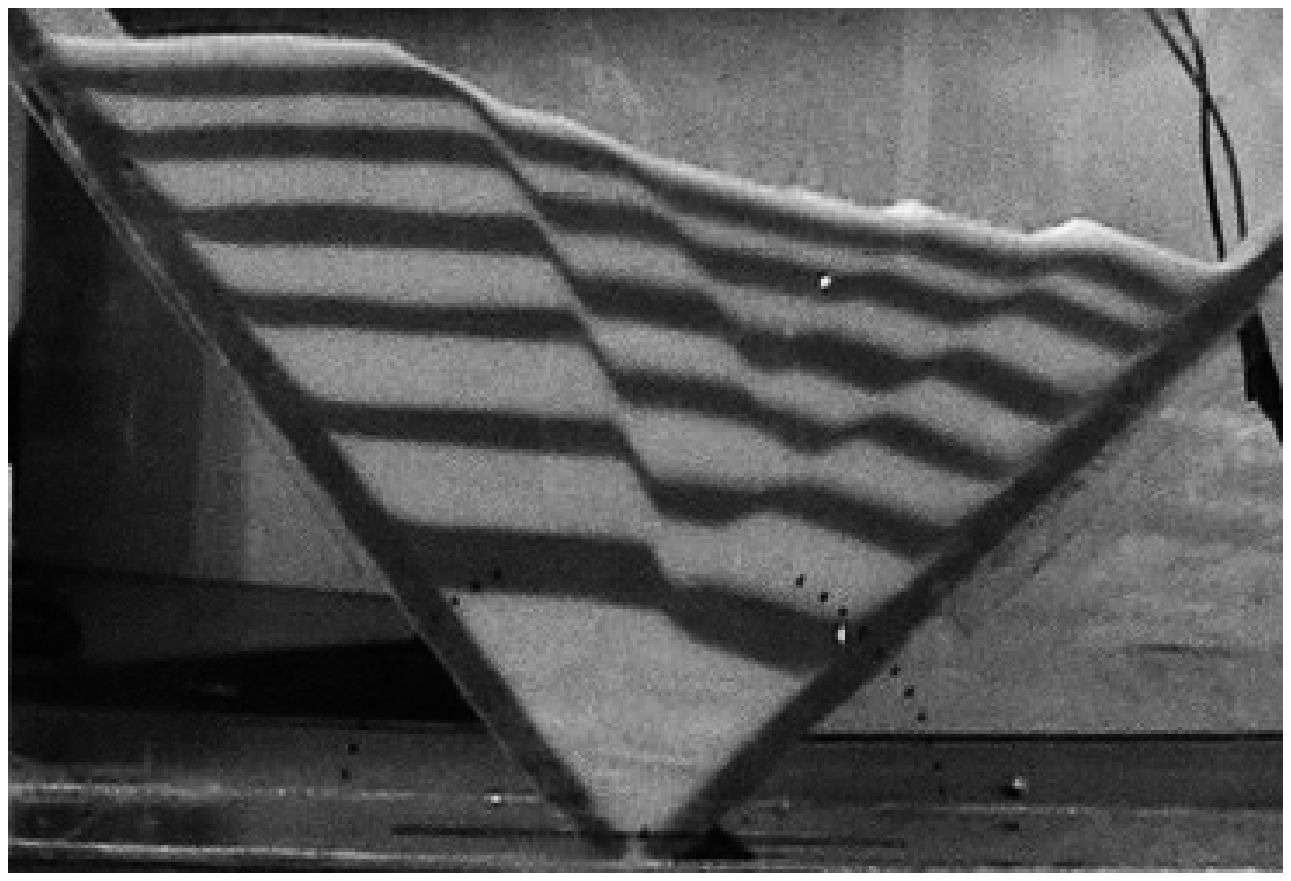}
\caption{A lateral view of the container at $\theta=45^{\circ}$ in the case that horizontal colored layers were prepared initially.
}\label{side}
\end{minipage}
\end{center}
\end{figure*}

We consider the case with GB100 beads and with parameter values $M=400g$,
$\theta=30^{\circ}$, $\theta_0=60^{\circ}$,
$\omega=0.60\pm0.05^{\circ}/s$ and $\theta_{fix}=50^{\circ}$ as the 
`standard'.
In the results described in this paper, we give explicit parameter values
 only when they differ from these.
In the actual experiments, the above-described setup was placed into an
airtight container.
Before carrying out the experiments, we dried the beads by placing them
in this container and using dessicants to keep the relative humidity in
the range $5-15\%$. 
This was done for several days, until their total weight no longer
decreased with time.

Lighting the surface obliquely to make the unevenness clearly visible, we
photographed the system from a position directly above, using a digital video
recorder.
Figure \ref{snapshot} displays a typical gray-scale image of a stripe pattern,  
where the width of the image corresponds to a length of $128.3mm$ on
the surface, and the vertical axis $y$ is parallel to the rotation axis of V-walls.
It is seen that stripes extend along the entire length of the granular
surface, except near the lateral walls, and that they are almost parallel to the $y$ axis.
We also see that the spacing of the stripes depends significantly on the
value of the $x$ coordinate.
In order to calculate the spacing of the stripes $\lambda(x,y)$ at the
place $(x,y)$ as a function of position, we processed a image as follows.

We cropped the central region of a snapshot, as shown in
Fig.\ \ref{snapshot} with dotted lines, and 
measured the brightness $b_{ij}$ at a set of points $\{(i,j)\}$ of the cropped digital image,
 where $x=i{\mathit \Delta} x$, $y=j{\mathit \Delta}x$ and ${\mathit \Delta}x=0.513mm$.
We first defined the left boundary as $i=l(j)$ and the right boundary as
$i=r(j)$ of the granular surface between V-walls 
by setting some threshold for $\langle b_{ij}\rangle$, 
where $\langle\cdot\rangle$ represents an average over the rectangular
 region of size $5\times 31$ pixels centered at $(i,j)$.
The small white rectangle in Fig.\ \ref{snapshot} displays this size. 
We next calculated the gradient of $b_{ij}$ with respect to $i$ as    
$s_{ij}\equiv \left(\langle ib_{ij}\rangle -\langle i\rangle \langle
b_{ij}\rangle \right)/ \left(\langle i^2 \rangle - \langle i\rangle^2
\right)$, using a least square method.
For each value of $j$, we determined the intervals in
which $s_{ij}$ takes negative values, and    
for the $k$th such interval of decreasing brightness, $I_{j}^{(k)}$, we calculated the `increment of
brightness' ${\mathit \Delta} s_{j}^{(k)}\equiv \sum_{i\in
I_{j}^{(k)}}s_{ij}$ and the average of $s_{ij}$, $\overline{s}_{j}^{(k)}$. 
We defined the right edge of any interval
$I_{j}^{(k)}$ that satisfies both conditions ${\mathit \Delta} s_{j}^{(k)}<-0.005$
and $\overline{s}_{j}^{(k)}<-0.0002$ to be the center of a dark stripe.
The value $0.005$ is on the order of the smallest difference in brightness
that can be distinguished from the digital images.

We denote to the centers of dark stripes as $i=c_j^{(m)}$, where
$m=1,2,...,m_j$ and $c_j^{m}<c_j^{(m+1)}$,  
and the boundaries of the granular surface as $c_j^{(0)}\equiv l(j)$ and
$c_j^{(m_j+1)}\equiv r(j)$. 
For given $y={\mathit \Delta}x j$, the spacing of stripes $\lambda(x,y)$ is
defined as the distance ${\mathit \Delta}x(c_j^{(m+1)}-c_j^{(m)})$, where
${\mathit \Delta}x c_j^{(m)}\leq x < {\mathit \Delta}x c_j^{(m+1)}$.
Using $\lambda(x,y)$ defined in this manner, at each $x$ we determined
its average over $y$ and over five experiments, $\overline{\lambda}(x)$,
and the corresponding standard deviation, ${\mathit \Delta}\lambda (x)$.
We display these functions in Fig.\ \ref{move}, where the thick curves
represent $\overline{\lambda}(x)$, with the ordinate scale appearing on
the left and the thin curves represent ${\mathit
\Delta}\lambda(x)/\overline{\lambda}(x)$, with the ordinate scale
appearing on the right.
We find that the quantity ${\mathit \Delta}\lambda/\overline{\lambda}$
is in the range $0.1-0.3$ for uniformly aligned stripes, while it takes larger values for disordered or branching stripes. 
Above the graph appear parts of snapshots cropped from the central
region with respect to $y$. 
The $x$ coordinates correspond to the abscissa of the graph below.

\

%%%%%%%%%%%%%%%%%%%%%%%%%%%%%%%%%%%%%%%%%%%%%%%%%%%%%%%%%%%%%%%%%%%%%%%%%%%
%\section{Results}
%%%%%%%%%%%%%%%%%%%%%%%%%%%%%%%%%%%%%%%%%%%%%%%%%%%%%%%%%%%%%%%%%%%%%%%%%%%
\begin{figure*}[t]
\begin{minipage}{5cm}
\epsfig{width=5cm,file=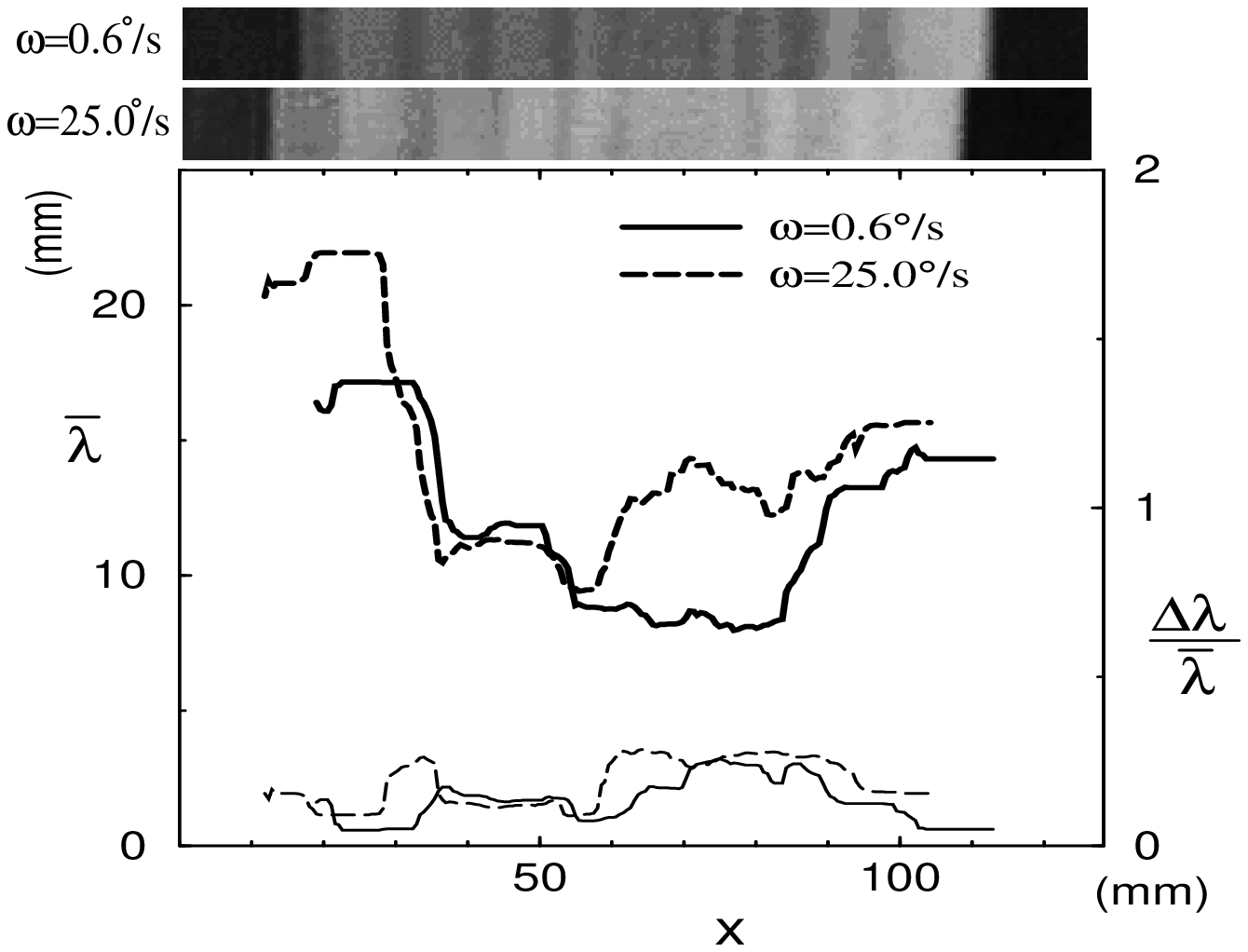}
\caption{Dependence on the angular velocity.
}\label{speed}
\end{minipage}\hspace{5mm}\
\begin{minipage}{5cm}
\epsfig{width=5cm,file=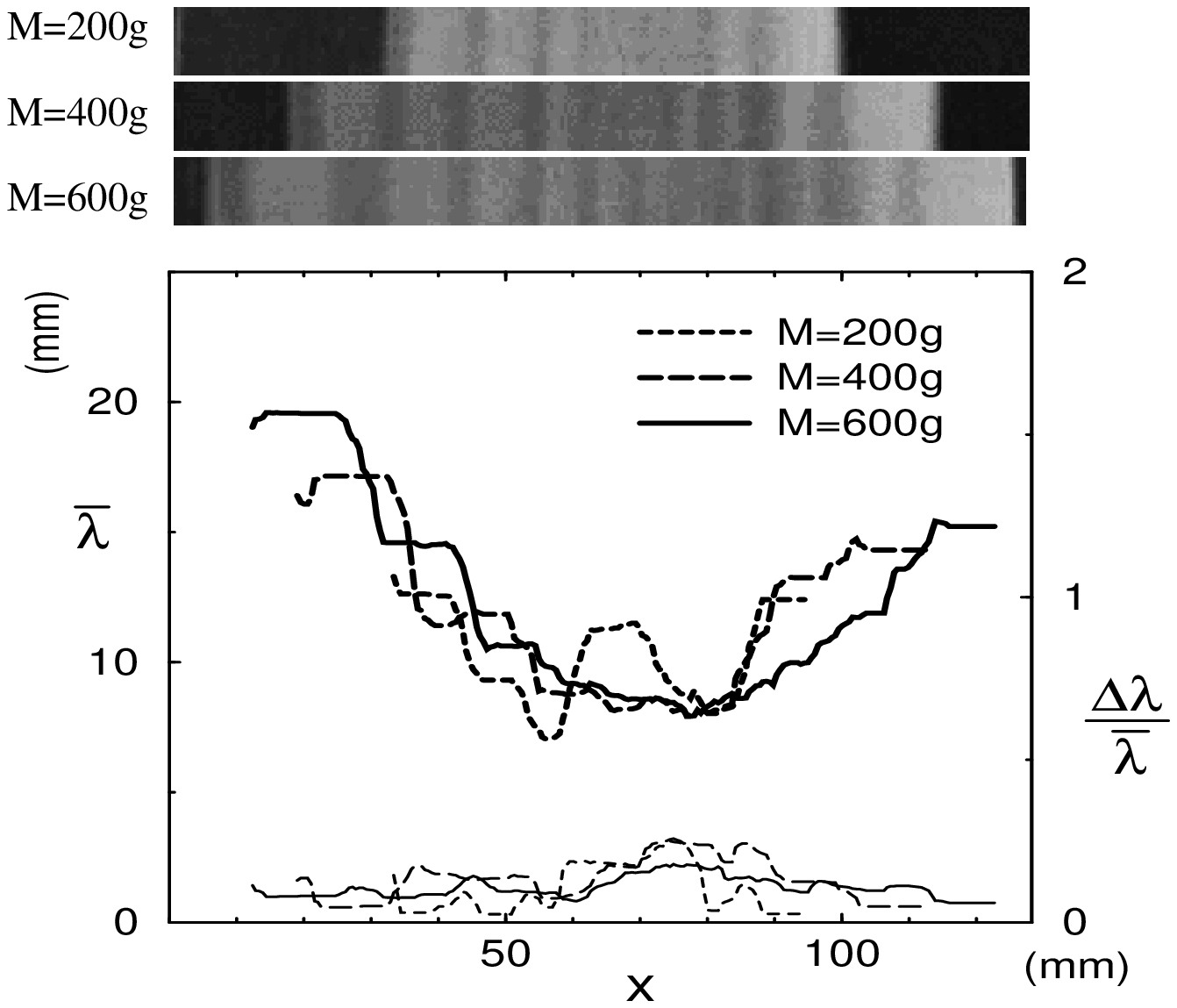}
\caption{Dependence on the system size.}\label{weight}
\end{minipage}\hspace{5mm}\
\begin{minipage}{5cm}
\epsfig{width=5cm,file=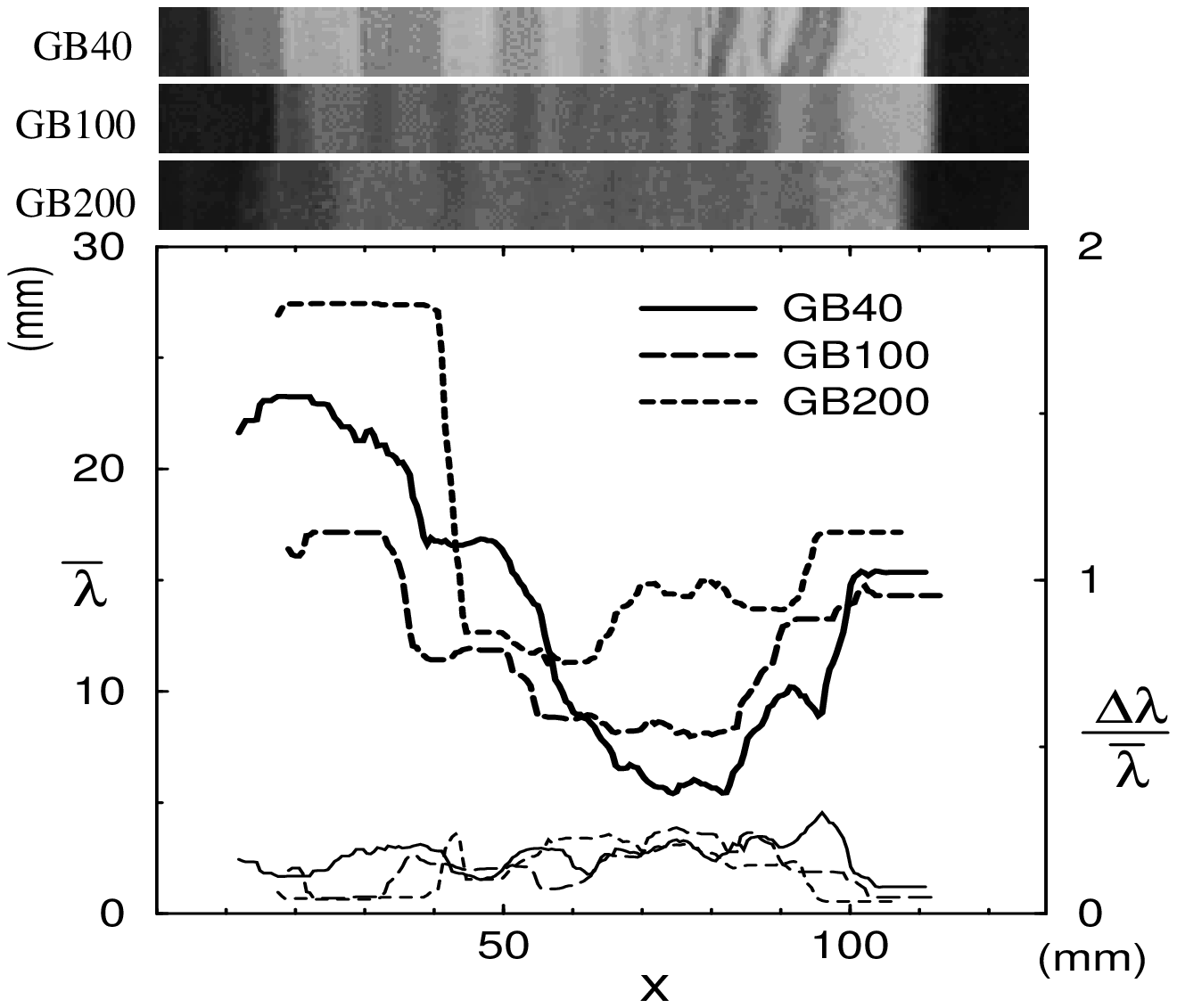}
\caption{Dependence on the grain size.}\label{radius}
\end{minipage}
\end{figure*}

We describe the experimental results below.

Figure \ref{move} displays a series of snapshots taken
 during the formation of a stripe pattern from $\theta_0=60^{\circ}$ to
 $\theta=30^{\circ}$ at decrements of $5^{\circ}$, and data for
 $\theta=50^{\circ},45^{\circ}$ and $40^{\circ}$. 
We find that stripes begin to form before $\theta_0-\theta$ reaches
 $10^{\circ}$, and their amplitudes increase as this value increases.
We also find that stripes near centered values of $x$ are observed later
 than those near V-walls.
Also, the spacing of stripes decreases when moving from either V-wall forward
 the center, along the $x$ axis.

We were able to observe the movement of grains at the lateral walls 
because the lateral walls are made of transparent acrylic.
In order to visualize the displacement of grains, 
we prepared horizontal layers of different colored beads prior to the
rotation of the plate.
In this case we used a larger container, with a setup essentially the
same as that described in Fig.\ \ref{setup}.
In this set of experiments, we used the glass beads GB-01 with $M=1000g$.
The beads were stained with a water-soluble ink.
According to our comparison of systems of stained and unstained beads,
there is no difference between the behavior of the two.
Figure \ref{side} displays a lateral view of the container.
Several faults are observed on each side of the V-shaped container.
We find that each fault appears as a stripe on the surface.
Also, apparently, every sufficiently thick stripe corresponds to some fault, 
although it was not possible to detect faults corresponding to narrow
stripes appearing near the central region with respect to $x$. 
We believe that this is due to disturbance of the lateral boundaries.
The faults we observed were found to be nearly straight, although long
faults exhibited slight curves.  
In our experiment, the most distinct fault, which appeared near the
fixed plate, made an angle of $63\pm 2^{\circ}$ with respect to the
horizontal. 
We note that the fault was not parallel to the fixed plate with $\theta_{fix}=50^{\circ}$бе 
While faults near the moving plate were formed at approximately the
same angle (inclined on the opossite direction), their slopes decreased as the plate was rotated.

When we set the angle of the fixed plate $\theta_{fix}$ at smaller
 values, we found that no stripe was formed in the region near the fixed plate, where the surface remains almost horizontal.
While the size of this region increased as $\theta_{fix}$ decreased,
 it was found that neither the width of the region in which stripes
 appear nor the spacing of stripes depends on $\theta_{fix}$.
These independences suggest that the inclination of the faults we 
 measured above was determined independently on $\theta_{fix}$.

We also conducted experiments in an air environment with higher
humidity($\sim 50\%$) and found that striped patterns appear in this
case too.
To test the case of very low humidity, we also carried out experiments in a
nitrogen environment.
Before the experiments, we filled the airtight container with nitrogen
and then flushed it several times, until the weight of the glass beads no longer decreased.
In comparing the results obtained in this case and that described in
Fig.\ \ref{move}, we were able to find no difference in the striped patterns appearing in each, within experimental error.
In the experiments of Terada and Miyabe mentioned above, they
investigated a granular system consisting of baked sand.
Our and their results together provide evidence that stripe patterns may
appear generally in completely dried granular systems.

We display the results for significantly different angular velocities
 $\omega=0.6$ and $25.0^{\circ}/s$ in Fig.\ \ref{speed}. 
We find that the spacing of stripes changes little as $\omega$ is
 increased, although the stripes become somewhat less pronounced.
We found that the case in which $\omega=0.6^{\circ}/s$ is essentially
 quasi-static.
Thus our results show that stripe patterns appear in the quasi-static
 deformation of a container and that the spacing is essentially independent on the deformation speed.

Figure \ref{weight} displays the data for experiments with different
 total weights of beads, $M=200, 400$ and $600$, for which the dimension of the system along the $x$ direction increases in proportion to $\sqrt{M}$.
The three functions $\overline{\lambda}(x)$ are identical, within
 experimental error.
We find that the spacing of stripes is almost independent of the system size.

In Figure \ref{radius}, we display the results for the three kinds of
glass beads GB40, GB100 and GB200.
In the case of GB40, the boundaries
 between the stripes are sharp, and we observe fine stripes in the
 central region.
By contrast, stripes in the central region are difficult to discern in
the case of GB200, which consists of grains with 5 times larger diameter
than GB40. 
Except in the central region, we found that the spacing of stripes does not depend significantly on the size of grains. 
It is, however, difficult to determine the exact dependence on the grain size 
because the size distributions and the initial volume fractions
of beads in the container differ slightly for the different kinds of beads.

\

%%%%%%%%%%%%%%%%%%%%%%%%%%%%%%%%%%%%%%%%%%%%%%%%%%%%%%%%%%%%%%%%%%%%%%%%%%%
%\section{Discussions}
%%%%%%%%%%%%%%%%%%%%%%%%%%%%%%%%%%%%%%%%%%%%%%%%%%%%%%%%%%%%%%%%%%%%%%%%%%%
As described above, the stripe patterns observed in our experiments
result from the formation of faults developing throughout the bulk of the system, not from surface flow.
It is well known that a typical shear band in a granular system has a
thickness on the order of several grain diameters, and it is thought to increase with the shear velocity \cite{deGennes98,Astrom00,Lemieux00}.
We believe that these dependences are the reason that the stripes become
less pronounced as either the size of the beads or the rate of
deformation of the container increases.

It is believed that a shear band in a dry granular system created in the plane in which the Mohr-Coulomb criterion 
$\sigma_t\geq \tan{\phi_c}\!\ \sigma_n \label{M.C.} $
 is first satisfied, where $\sigma_n$ and $\sigma_t$ represent the normal and
 shear components of the stress on this plane.
The constant $\phi_c$ is called the `maximum angle of internal friction'.
By regarding the system as a two-dimensional continuous medium and
assuming that this criterion holds critically
 [i.e. $\sigma_t=\tan{\phi_c}\!\ \sigma_n$] at a fault, the inclination
 of the faults $\phi$ is given by $\phi=\phi_c/2+45^{\circ}$ with respect to the minimum principal axis of the stress \cite{deGennes98,Astrom00,Terada}.
In our system, the above-described results show that the faults formed 
near the beginning of the rotation, at a time when the granular surface
was still nearly horizontal. 
If we assume that the minimum principal axis was parallel to this
granular surface, we can regard the value of $\phi$ given above to be measured
with respect to the horizontal.
$\phi_c$ is usually considered, as an approximation, the maximum angle 
of the slope at which there occurs no surface flow. 
Measuring this maximum angle for GB-01 to be $28.1\pm 0.7^{\circ}$, 
we estimate that the angle of the faults should be $\phi=59.0\pm
0.4^{\circ}$.
While this is not consistent with the experimental value $63\pm
2^{\circ}$, these values are quite close, and thus it appears that the 
assumptions made above are reasonable as the first approximation. 

As stated above, the spacing of stripes does not depend
significantly on either the system size or the grain size,
 and the phenomena we have studied are induced by quasi-static deformation.
In dry granular systems consisting of spherical rigid particles, 
frictional coefficients and the acceleration due to gravity are the only 
parameters characterizing the forces involved in the quasi-static dynamics. 
A simple dimensional analysis shows that such systems have no characteristic length other than the system size and the grain size.
For this reason, we believe that the mechanism responsible for arranging
faults possesses no characteristic length to determine their spacing.
We note that this differs from the situation involving shear bands appearing in rocks \cite{Poliakov94}.  
Our finding that the spacing of stripes becomes smaller as $x$
approaches the central region leads us to believe that the spacing in a
particular region depends on the depth of the faults in this region that 
create the stripes.

\

%%%%%%%%%%%%%%%%%%%%%%%%%%%%%%%%%%%%%%%%%%%%%%%%%%%%%%%%%%%%%%%%%%%%%%%%%%%
%\section{Conclusions}
%%%%%%%%%%%%%%%%%%%%%%%%%%%%%%%%%%%%%%%%%%%%%%%%%%%%%%%%%%%%%%%%%%%%%%%%%%%

In summary, we have reported experimental results for a granular
system in which stripe patterns result from the slow 
deformation of a V-shaped container. 
Our results show that this behavior occur both in the dry granular
systems and in quasi-static deformation.
The stripe patterns are caused by the formation of many
faults, which are arranged approximately parallel to one another on each
side of the V-shaped container.
From the measurement of the inclination of the faults, we conclude that 
the stress on each faults is approximately critical at the time that the
stripes begin to form.
We also found that the spacing of stripes becomes smaller in the central
region, away from V-walls, and depends significantly on
neither the system size nor the grain size. 
These results suggest that the mechanism responsible for arranging the
faults possesses no characteristic length to determine the spacing.

%%%%%%%%%%%%%%%%%%%%%%%%%%%%%%%%%%%%%%%%%%%%%%%%%%%%%%%%%%%%%%%%%%%%%%%%%%%
%\section{Acknowledgements}
%%%%%%%%%%%%%%%%%%%%%%%%%%%%%%%%%%%%%%%%%%%%%%%%%%%%%%%%%%%%%%%%%%%%%%%%%%%
We acknowledge C. Urabe, N. Yazima and
H. Hayamawa for fruitful discussions, G.C. Paquette for reading the
manuscript conscientiously, and K. Yamamoto and J. Karimata 
for providing experimental apparatus and technical advice.
We thank S. Nasuno, M. Toda and the members of the Nakaya Ukichiro Museum of Snow and Ice for providing information on the works of T. Terada and N. Miyabe.

%%%%%%%%%%%%%%%%%%%%%%%%%%%%%%%%%%%%%%%%%%%%%%%%%%%%%%%%%%%%%%%%%%%%%%%%%%%
%\section{References}
%%%%%%%%%%%%%%%%%%%%%%%%%%%%%%%%%%%%%%%%%%%%%%%%%%%%%%%%%%%%%%%%%%%%%%%%%%%

\end{document}